\documentclass[%
 reprint,
 superscriptaddress,
 amsmath,amssymb,
 aps,
pra,
floatfix,
10pt,
]{revtex4-2}

\usepackage{graphicx}
\usepackage{dcolumn}
\usepackage{bm}
\usepackage[utf8]{inputenc}
\usepackage[T1]{fontenc}
\usepackage[english]{babel}
\usepackage{floatrow}
\usepackage[labelformat=simple, caption=false]{subfig}
\usepackage[input-uncertainty-signs]{siunitx}
\usepackage{hyperref}
\usepackage{cleveref}
\usepackage{stackengine}
\usepackage[export]{adjustbox}
\usepackage{nicefrac}
\usepackage{xcolor}
\usepackage[nolist,nohyperlinks]{acronym}
\usepackage{float}
\usepackage{placeins} 

\newcommand\Membranethickness{850}
\newcommand\Membranethicknesstwo{3}
\newcommand\Membranethicknessimplant{1.2}
\newcommand\Membraneroughness{0.44}
\newcommand\Implantationdepth{50}

\newcommand\yieldcreation{5}
\newcommand\emitterdensity{1.5}
\newcommand\numberemitter{85}
\newcommand\NumberEmitterConfocal{45} 
\newcommand\NumberEmittersCav{150} 
\newcommand\Emitterlinewidthsingle{171.4}
\newcommand\LinewidthinhomoFSfourK{25.1} 
\newcommand\LinewidthinhomoCavfourK{65.7} 
\newcommand\LifetimeEmitterFS{1.74} 
\newcommand\LifetimeEmitterFSSeventyK{1.71}
\newcommand\LifetimeEmitterFSRT{1.2}
\newcommand\groundstatesplitmin{170} 
\newcommand\groundstatesplitmax{550} 
\newcommand\excitedstatesplitmin{380} 

\newcommand\ROCx{17.6} 
\newcommand\ROCy{21.7} 
\newcommand\ROC{19.7}
\newcommand\ellipticity{0.19}
\newcommand\CavityLength{10} 
\newcommand\CavityLinewidth{5.4}
\newcommand\ModeVolume{45.2} 
\newcommand\ModeNumber{27}
\newcommand\MirrorReflectivity{99.9}
\newcommand\MirrorTransppm{1000}
\newcommand\finesseTheo{3140}
\newcommand\finesseExp{2800}
\newcommand\QualityFactor{76000}
\newcommand\BeamWaist{1.52}
\newcommand\CavityModeArea{1.81\,}

\newcommand\PurcellExp{2.2} 
\newcommand\PurcellTheo{9.2} 
\newcommand\LifetimeEmitterCavity{0.8} 
\newcommand\LifetimeEmitterCavitySev{1.44} 
\newcommand\Cooperativity{1.2}
\newcommand\betafactor{0.54}
\newcommand\gcavity{0.43}
\newcommand\gammacavity{0.091}

\newcommand\LossesMaxreduction{2710} %
\newcommand\sigmamaxreduction{4.9\,  \cdot 10^{-11}} %
\newcommand\sigmamaxreductionsingle{3.2\,  \cdot 10^{-13}} %
\newcommand\sigmawarmens{1.3} %
\newcommand\sigmamiddleens{1.6} %
\newcommand\sigmacoldens{1.9} %

\newcommand\RefractionIndex{2.4} 
\newcommand\straintransversemax{2\, \cdot 10^{-4}} %

\newcommand\powerPLEsingle{450}

\begin{document}
\title{Cavity-Enhanced Emission and Absorption of Color Centers in a Diamond Membrane With Selectable Strain} %

\author{Robert Berghaus}%
\author{Selene Sachero}
\author{Gregor Bayer}
\affiliation{%
 Institute for Quantum Optics, Ulm University, Albert-Einstein-Allee 11, 89081 Ulm, Germany}%
\author{Julia Heupel}
\affiliation{%
 Institute of Nanostructure Technologies and Analytics, Center for Interdisciplinary Nanostructure Science and Technology, University of Kassel, Heinrich-Plett-Str. 40, 34132 Kassel, Germany}%
\author{Tobias Herzig}
\affiliation{%
Division of Applied Quantum Systems, Felix Bloch Institute for Solid State Physics, University Leipzig, Linnestraße 5, 04103 Leipzig, Germany}%
\author{Florian Feuchtmayr}
\affiliation{%
 Institute for Quantum Optics, Ulm University, Albert-Einstein-Allee 11, 89081 Ulm, Germany}%
\author{Jan Meijer}
\affiliation{%
 Division of Applied Quantum Systems, Felix Bloch Institute for Solid State Physics, University Leipzig, Linnestraße 5, 04103 Leipzig, Germany}%
\author{Cyril Popov}
\affiliation{%
 Institute of Nanostructure Technologies and Analytics, Center for Interdisciplinary Nanostructure Science and Technology, University of Kassel, Heinrich-Plett-Str. 40, 34132 Kassel, Germany}%
\author{Alexander Kubanek}
\thanks{Corresponding author}
\email{alexander.kubanek@uni-ulm.de}
\affiliation{%
 Institute for Quantum Optics, Ulm University, Albert-Einstein-Allee 11, 89081 Ulm, Germany}%

\begin{abstract}

Group IV color centers in diamond are among the most promising optically active spin systems with strong optical transitions and long spin coherences. The ground-state splitting of the center is particularly important to suppress the interaction with coherence-limiting phonons, which improves the coherence properties and sets the upper limit for the operating temperature. Negatively charged silicon-vacancy centers have an ordinary ground-state splitting of only $\SI{48}{\giga\hertz}$, resulting in required temperatures below one Kelvin, which can only be achieved by dilution refrigerators. Here, we increase the ground-state splitting by up to an order of magnitude by induced strain in a single-crystal diamond membrane. Furthermore, we demonstrate cavity-assisted spectroscopy enabled by coupling the emitter ensemble with a selectable strain to the mode of a Fabry-Perot microcavity. Calculation of the absorption cross-section yields $\sigma_{\mathrm{abs}}^{\mathrm{ens}}=\sigmamaxreduction\,\mathrm{cm}^2$. Together with the Purcell-enhanced twofold reduction in emitter lifetime below $\SI{1}{\nano \second}$, this makes the system a promising spin-photon interface at moderate temperatures of $\SI{4}{\kelvin}$.

\end{abstract}

\maketitle
\begin{acronym}[Basht]
\acro{ZPL}{zero phonon line}
\acro{FP}{Fabry-Perot}
\acro{DM}{diamond membrane}
\acro{DBR}{distributed Bragg reflector}
\acro{PL}{photoluminescence}
\acro{PLE}{photoluminescence excitation}
\acro{RoC}{radius of curvature}
\end{acronym}
\renewcommand{\figurename}{Fig.}
\renewcommand{\tablename}{Tab.}
\sisetup{uncertainty-mode = compact-marker}


\section*{Introduction}

Diamond, known as the hardest natural material, offers many outstanding properties not only for the classical world but also for fields such as quantum optics. A wide electronic band gap enables optical transparency and, at the same time, opens the possibility to host many different quantum emitters, known as color centers. In addition, the low concentration of intrinsic nuclear spins within the carbon lattice is an important prerequisite for isolating individual spins and avoiding unwanted interactions with the environment. Group IV color centers in diamond possess a strong optical transition with down to lifetime-limited linewidths at liquid helium temperatures. Their electron and nuclear spins are optically accessible, offering millisecond-long coherence times \cite{metsch_initialization_2019}. The electron spin coherence time is limited by a direct phonon process coupling both ground states \cite{jahnke_electronphonon_2015}. Thus, a larger ground state splitting yields a longer electron spin coherence time \cite{iwasaki_tin-vacancy_2017,ruf_quantum_2021}. 
Group IV color centers in unstrained diamond crystal display an increasing ground-state splitting with increasing atomic number of the interstitial atom. Thus, those with higher atomic numbers are associated with longer electron-spin coherence times at moderate temperatures above 1\,Kelvin. However, group IV centers with lighter atoms, such as the negatively-charged silicon-vacancy (SiV$^{\text{-}}$) center, can be engineered via strain to create a large ground-state splitting \cite{meesala_strain_2018, sohn_controlling_2018, assumpcao_deterministic_2023}.
The effect of strain has been studied for different color centers in diamond.\cite{nguyen_integrated_2019,guo_microwave-based_2023, karapatzakis_microwave_2024}. 
While strain control is a promising way to improve the spin coherence properties, an efficient interface to photons is required to establish a spin-photon interface. Therefore, the coupling of the optical transition of the color centers to a predefined optical mode can be optimized through integration into an optical cavity \cite{janitz_cavity_2020}. \ac{FP} resonators benefit from their full tunability, which makes them versatile. For example, the mode's resonance frequency can be adjusted to the transition frequency of the color center, which can be highly shifted by strain.
While the successful integration of diamond into \ac{FP} resonators has been demonstrated in numerous experiments \cite{kaupp_scaling_2013, albrecht_coupling_2013, benedikter_cavity-enhanced_2017,hoy_jensen_cavity-enhanced_2020, ruf_resonant_2021, flagan_diamond-confined_2022, feuchtmayr_enhanced_2023, heupel_fabrication_2023,korber_scanning_2023,yurgens_cavity-assisted_2024}, cavity enhancement of group IV centers in \ac{FP} resonators at cryogenic temperatures remains challenging. It yet has only been achieved in a small number of experiments as shown with the SiV$^{\text{-}}$ center \cite{salz_cryogenic_2020,bayer_optical_2023} as well as with the negatively-charged germanium-vacancy (GeV$^{\text{-}}$) center \cite{zifkin_lifetime_2024} and tin-vacancy (SnV$^{\text{-}}$) center \cite{herrmann_coherent_2023}. 
Diamond integration is achieved using either a smooth \ac{DM} or a nano-particle for low scattering losses and short cavity lengths.

Here, we couple a \ac{DM} with implanted SiV$^{\text{-}}$ color centers under strain to the mode of an open-access \ac{FP} microcavity. 
We fabricated the single-crystal \ac{DM} with ensembles of SiV$^{\text{-}}$ centers and show how different strain regions shift the SiV$^{\text{-}}$ optical transitions allowing for increased ground-state splitting as exploited in GeV$^{\text{-}}$ and SnV$^{\text{-}}$. This allows for the selective use of emitters with a different and intrinsic strain environment.
By integrating the \ac{DM} with negligible scattering losses into our tunable \ac{FP} cavity, we reach finesse values of up to $\finesseExp$ and address the ensemble with cavity-assisted \ac{PL} and \ac{PLE}. We enhance the optical C-transition of the SiV$^{\text{-}}$ centers with the cavity-emitter system due to the twofold Purcell effect and establish absorption spectroscopy within our cavity, where we measure an absorption cross-section of the emitter ensemble of up to $\sigma_{\mathrm{abs,ens}}=\sigmamaxreduction\,\mathrm{cm}^2$.

\begin{figure*}[ht]
  \centering
   \includegraphics[scale=1]{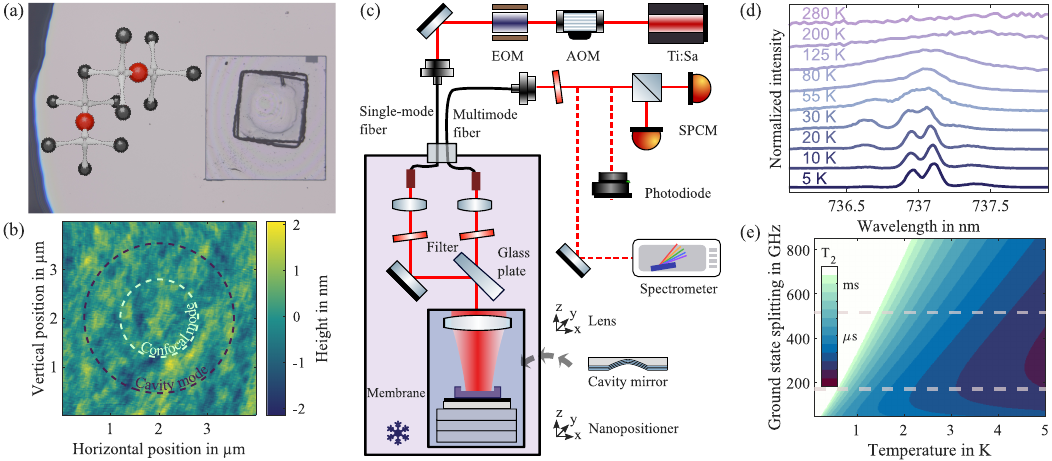}
    \caption{\textbf{
        A strained single-crystal diamond membrane with implanted SiV$^{\text{-}}$ centers for integration into an open-access microcavity.}
\textbf{(a)} Microscope image of the thin diamond membrane bonded to the flat cavity mirror. The two overlapping rectangles indicate the etched inner part of the membrane. Inside: Sketch of the diamond lattice showing two different SiV$^{\text{-}}$ centers. 
\textbf{(b)} Atomic force microscope image of the etched membrane after fabrication, representing a low surface roughness of below $\SI{0.5}{\nano \meter}$, with the confocal and cavity mode areas indicated. 
\textbf{(c)} Sketch of the cryogenic experimental setup for confocal microscopy and cavity experiments, with the upper concave mirror mounted. The purple shaded area indicates the bath cryostat and the blue shaded area the confocal/cavity part.
\textbf{(d)} Temperature-dependent photoluminescence spectra of the SiV$^{\text{-}}$ emitter ensemble in confocal mode, indicating two classes of emitters.
\textbf{(e)} Expected order of the coherence time $T_2$ as a function of different ground-state splittings $\Delta_{\mathrm{gs}}$ and temperature due to phonon limitations.
}
\label{fig1}
\end{figure*}

\section*{Methods}

\subsection*{Single-Crystal Diamond Membrane}

The single-crystal \ac{DM} was fabricated by inductively coupled plasma reactive ion etching (ICP-RIE; Plasmalab 100, Oxford Instruments plc), according to reference \cite{heupel_fabrication_2020}. Prior to structuring, the electronic grade single-crystal diamond ($\SI{20}{\micro\meter} \times \SI{2}{\milli\meter} \times \SI{2}{\milli\meter}$; Applied Diamonds Inc.) was cleaned with piranha solution (3:1, 1\,h) and treated in an oxygen plasma asher ($\SI{150}{\watt}$, $\SI{0.7}{\milli\bar}$, 2\,min; TePla 200-G, PVA TePla AG). To withstand the long etching processes and to create a thinned diamond area of  $\SI{1}{\milli\meter} \times \SI{1}{\milli\meter}$, a rectangular bulk diamond mask (Medidia GmbH) was utilized. The mask provides an angled inlet to prevent over-etching towards the membrane edges. 
Using an Ar/Cl$_2\ + \ $O$_2$ cyclic recipe, the diamond area was thinned to a minimum thickness of $\SI{\Membranethickness}{\nano\meter}$, see \hyperref[sec:appendix:thickness]{Appendix \ref{sec:appendix:thickness}}. The Ar/Cl$_2$ steps provide a smoothing effect, while the O$_2$ steps have a high etch rate to etch deep into the diamond.

\subsection*{Creation of Color Centers}

A dense SiV$^{\text{-}}$ ensemble (fluence of $3\cdot10^{11}$ per cm$^2$) created by high dose implantation of $^{28}$Si at $\SI{72}{\kilo \electronvolt}$ resulted in a mean implantation depth of $\SI{\Implantationdepth}{\nano \meter}$. 
95\,\% of the ions reached a depth within the \ac{DM} between $\SI{25}{\nano \meter}$ and $\SI{75}{\nano \meter}$, see \hyperref[sec:appendix:fab]{Appendix \ref{sec:appendix:fab}}.
Implantation was followed by high-temperature annealing \cite{lang_long_2020}, for 1\,h at $1200 ^{\circ}$C and $1500 ^{\circ}$C and cleansing with hydrofluoric acid and tri-acid.
From the fluence and an expected emitter creation yield of $\SI{\yieldcreation}{\percent}$ per silicon ion we estimated an emitter density of $\emitterdensity\cdot10^{10}$ per cm$^2$. 

\subsection*{Experimental Platform}

\hyperref[fig1]{Fig.\,1(a)} displays the implanted single-crystal \ac{DM} bonded to a flat \ac{DBR} for later cavity integration. 
For a <100> diamond facet, the projected dipoles of four possible SiV$^{\text{-}}$ orientations within the $\mathrm{D}_{3\mathrm{d}}$ symmetry form two orthogonal classes \cite{meesala_strain_2018}.
The two classes are sketched with the red marked silicon atoms surrounded by carbon atoms overlaid with the microscope image. This results in a perpendicular dipole emission for all four possible transitions between the two emitter classes. 
For cavity integration, a smooth diamond surface is necessary to minimize scattering losses. Additionally, the even surface enables strong bonding to the mirror via van der Waals forces \cite{heupel_fabrication_2020}. 
An atomic force microscopy scan (\hyperref[fig1]{Fig.\,1(b)}) of the \ac{DM} surface reveals a planar surface with a root mean square roughness of less than $\SI{0.5}{\nano \meter}$ within the scan area. The confocal and cavity mode areas are shown for comparison. Resulting scattering losses below $160\,$ppm allow finesse values $\mathcal{F}$ above 40000.

\hyperref[fig1]{Fig.\,1(c)} depicts the optical confocal setup, with the flexibility to include the curved cavity mirror for cavity assembly within the same setup. Whether free space or cavity experiments were conducted within the liquid helium bath cryostat (Oxford Instruments plc), the setup was cooled indirectly by a liquid nitrogen shield, or immersed in liquid helium for low temperatures and narrow optical transitions.
The hemispherical cavity was formed by a macroscopic flat mirror and a curved mirror, structured by CO$_2$-laser ablation. For $\SI{737}{\nano \meter}$, the transmission of the \ac{DBR} coatings was designed and measured as $T=1000\,$ppm, allowing for a finesse of up to $\mathcal{F}_\mathrm{theo}=\finesseTheo$, with a stopband between $\SI{610}{\nano\meter}$ and $\SI{750}{\nano\meter}$. For longer wavelengths, the transmission then increased to around $T\approx \SI{1}{\percent}$ at $\SI{788}{\nano \meter}$. 
The plane cavity mirror was placed on nanopositioners (ANPx101 and ANPz101, attocube systems AG) to scan the \ac{DM} and for cavity length tunability. 
For excitation and detection an aspherical lens ($f=\SI{8}{\milli\meter}$, NA$=0.55$; AHL10-08-P-U, asphericon GmbH) focused into the diamond in confocal mode or coupled the optical microcavity when the upper curved mirror was placed inside the cryostat, see \hyperref[sec:appendix:cavity]{Appendix \ref{sec:appendix:cavity}} for more details.

Temperature-dependent spectra of the SiV$^{\text{-}}$ ensemble are shown in \hyperref[fig1]{Fig.\,1(d)}. The double peak, most prominent at \SI{5}{\kelvin}, originates from the two emitter classes. Strain shifts the transitions in the spectral domain due to different transverse and longitudinal strain components on the emitters.  While the transverse strain affects the ground-state splitting, the longitudinal strain mainly shifts the entire \ac{ZPL} spectrum.
An increased ground-state splitting increases the upper limits for the spin coherence time $T_2$, which is mainly limited by acoustic phonon processes \cite{jahnke_electronphonon_2015}. The expected order of the coherence time, based on one phonon interaction \cite{jahnke_electronphonon_2015}, is shown in  \hyperref[fig1]{Fig.\,1(e)} depending on the orbital ground-state splitting $\Delta_{\mathrm{gs}}$ and the operating temperature. 
Spin coherence control and a prolongation could be demonstrated for SiV$^{\text{-}}$ centers in nano-electrical-mechanical systems with externally applied strain \cite{sohn_controlling_2018,meesala_strain_2018}.

\begin{figure*}[ht]
  \centering
   \includegraphics[scale=1]{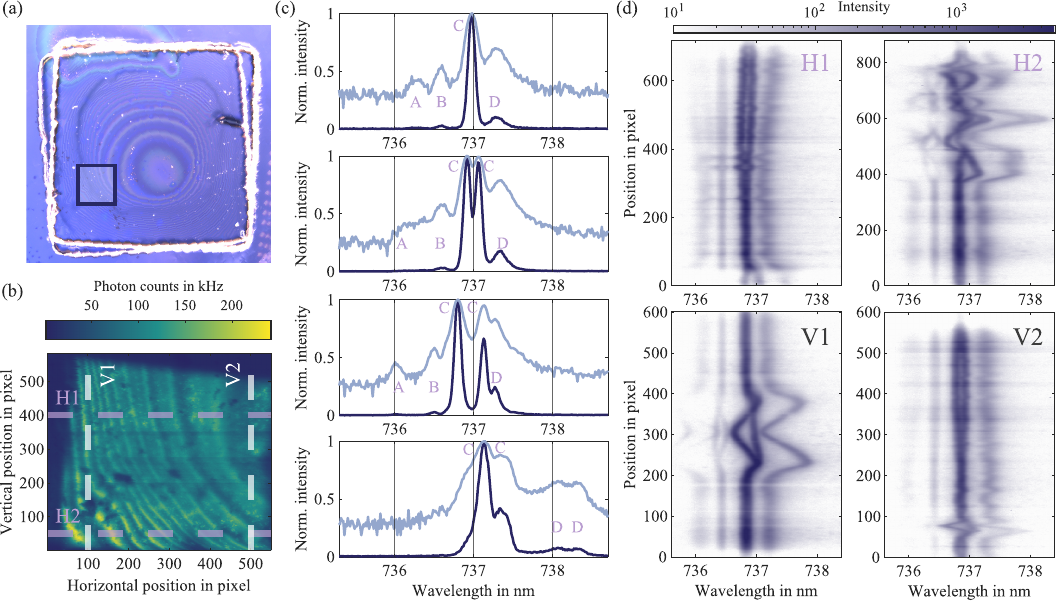}
    \caption{\textbf{
        Strain-dependent zero phonon line emission of the SiV$^{\text{-}}$ emitter ensemble at 4\,K.}
    \textbf{(a)} Differential interference contrast microscope image of the etched membrane. Fringes indicate height changes. The blue rectangle represents the implanted region.
    \textbf{(b)} Confocal membrane-on-mirror characterization of the implanted area under off-resonant excitation. Fringes result from decreasing thickness toward the upper right of the confocal scan. Dashed lines indicate line scans from the spectral analysis in (d).
    \textbf{(c)} Spectra (dark blue) showing different strain configurations. The spectra are also shown with a logarithmic scaling (light blue) for better visibility of the less prominent peaks.
    \textbf{(d)} Emission wavelengths of SiV$^{\text{-}}$ emitters along horizontal and vertical lines depicted in confocal scan (a). A spectrum was recorded at each position of the line scan with scan directions from left to right and bottom to top. The two bright lines in each scan show the C-transition of two different SiV$^{\text{-}}$ orientations. Spectral shifts are a result of non-uniform strain in the diamond crystal. 
    }
    \label{fig2}
\end{figure*}

\section*{Results and Discussion}

Spectroscopic analysis of the SiV$^{\text{-}}$ emitter ensemble was performed at $\SI{4}{\kelvin}$ directly on the mirror in confocal mode. The region of interest is marked in the microscope image \hyperref[fig2]{Fig.\,2(a)}, which corresponds to the densely implanted part. Thickness variations become evident by following the interference fringes \cite{Challier_advanced_2018} from the thin (\Membranethickness$\,$nm) middle part towards the thicker implanted region (up to \Membranethicknesstwo$\,\mu$m). 
Under off-resonant excitation around $\SI{712}{\nano \meter}$, photoluminescence of the \ac{ZPL} was collected.
While scanning the \ac{DM} on the mirror horizontally and vertically the count rate within the implanted region was recorded (\hyperref[fig2]{Fig.\,2(b)}). Fringes resulting from thickness variations follow a periodicity of $\lambda/(2n)$ \cite{korber_scanning_2023}, with $n=\RefractionIndex$ the refractive index of diamond.

\subsection*{Emitter Ensemble Strain Spectroscopy}

Selected spectra from differently strained regions are shown in \hyperref[fig2]{Fig.\,2(c)}. The typical 
SiV$^{\text{-}}$ ZPL emission  (upper panel) at $\SI{737}{\nano \meter}$ with a four-line structure, labeled A to D, was observed for the ensemble. 
The ground-state splitting of $\Delta_{\mathrm{gs}}=\SI{\groundstatesplitmin}{\giga \hertz }$ is more than three times larger than the spin-orbit coupling of unstrained SiV$^{\text{-}}$. The excited-state splitting with $\Delta_{\mathrm{es}}=\SI{\excitedstatesplitmin}{\giga \hertz }$ shows an increase of 1.5.
From the emitter density, we estimated an average number of $N\approx\numberemitter$ contributing emitters in the confocal mode $A=\frac{\pi}{4} \omega_{\mathrm{fs}}^2 $ with a beam waist of $\omega_{\mathrm{fs}}\approx \frac{2 \lambda}{\pi \mathrm{NA}} \approx \SI{0.8}{\micro\meter}$.
Resonant excitation of strongly axially strained emitters with transition wavelengths above $\SI{738}{\nano \meter}$, revealed single photon emission. 
See \hyperref[sec:appendix:single]{Appendix \ref{sec:appendix:single}} for the first report of a single SiV$^{\text{-}}$ inside a \ac{DM}.
For the two perpendicular emitter orientations, the strain environment shifts the transitions differently (\hyperref[fig2]{Fig.\,2(c)}, second and third panels). The most prominent C-line is therefore split. 
At a strong transverse strain of the order of $\straintransversemax$ with respect to \cite{meesala_strain_2018}, the ground states split by $\SI{\groundstatesplitmax}{\giga \hertz}$, corresponding to almost $\SI{1}{\nano\meter}$ (\hyperref[fig2]{Fig.\,2(c)}, lower panel). This corresponds to an 11-fold increase in ground-state splitting compared to no strain.

We spectrally map the strain of the emitter ensemble over a wide range within the single-crystal membrane.
Horizontal (H1, H2) and vertical line (V1, V2) scans with the changing \ac{ZPL} emission are shown in \hyperref[fig2]{Fig.\,2(d)}. 
The two emitter classes experience strong strain-induced shifts. 
Both classes are affected in regions of high strain, indicated by large shifts in the different optical transitions. Some spectral lines originating from different emitter classes feature crossings of spectral lines.
We attribute the strain to the bond, different thermal expansion coefficients of diamond and \ac{DBR}, and the fabrication process. A comparable ground-state splitting of up to  $\Delta_{\mathrm{gs}}\approx \SI{500}{\giga \hertz}$ was shown in a nanophotonic crystal cavity with implanted SiV$^{\text{-}}$ centers \cite{nguyen_integrated_2019}, where strain variations are attributed to the fabrication process. 
For the heavier group IV color centers including SnV$^{\text{-}}$, strain can simplify spin control due to required orbital mixing in the spin ground states \cite{Rosenthal_Microwave_2023,guo_microwave-based_2023,pieplow_efficient_2024}. Hence, an intentional mismatch of the thermal expansion coefficients between a diamond membrane and the substrate rendered a tensile strain on SnV$^{\text{-}}$ centers within the crystal in recent works \cite{guo_microwave-based_2023, karapatzakis_microwave_2024}.
Work on \ac{FP} cavity-enhanced GeV$^{\text{-}}$ reported an increase of the ground-state splitting up to a factor of 7 \cite{zifkin_lifetime_2024}. 
While these works probe the strain at a few points, we show the strain profile over a wide range, especially up to a high-strain regime.
Emitters with desired strain properties can be selected by adjusting the position of the membrane in the optical mode. In particular, these include emitters with a large ground-state splitting that possess superior coherence properties and promise operating temperatures above  $\SI{1}{\kelvin}$.

\subsection*{Cavity Integration}

\begin{figure*}[ht]
 \includegraphics{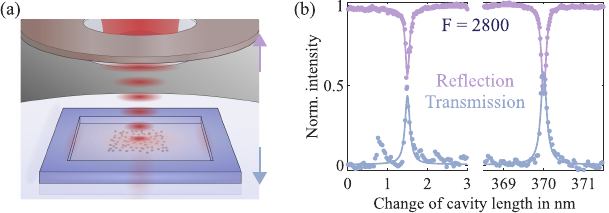}
  \centering
    \caption{\textbf{
        Cavity-integrated diamond membrane.}
    \textbf{(a)} Sketch of the membrane bonded to the flat mirror of the cavity, with the excitation laser coupled from above through the curved mirror. 
    \textbf{(b)} Cavity length scans over two cavity resonances in reflection (purple) and transmission (blue) at $\SI{80}{\kelvin}$. Data points as dots and Lorentzian fits as lines. The transmission signal was normalized to the reflection signal, without considering additional losses. 
    }
    \label{fig3}
\end{figure*}

We closed the cavity by mounting a concave mirror with a \ac{RoC} of $\SI{\ROC}{\micro \meter}$ and an ellipticity of \ellipticity. This allowed cavity-assisted experiments within the same setup. The schematic in \hyperref[fig3]{Fig.\,3(a)} depicts the mirror configuration and indicates the direction of the reflection and transmission signals from the microcavity.
Often, \ac{FP} microcavities are composed of a mirror-fiber combination \cite{ruf_resonant_2021, vadia_open-cavity_2021,hausler_tunable_2021} or even a fiber-fiber \cite{hunger_fiber_2010, ruelle_tunable_2022} assembly. For an efficient and mode-matched operation, we used two mirrors on plain silica substrates and coupled free space into fiber optics \cite{tomm_bright_2021}. 
The cavity length could be controlled by tuning the piezo crystal voltage of the z nanopositioner. 
X and y nanopositioners allowed the positioning of the plane mirror, here such that the top right of the implanted region lay within the cavity mode.

\hyperref[fig3]{Fig.\,3(b)} shows two typical cavity-length scans probing the resonance of a fundamental Gaussian mode on the diamond membrane. A laser applied to the cavity revealed the reflected and transmitted intensity, in this case at $\SI{80}{\kelvin}$.
The cavity length change was calculated by considering a length difference of $737/2\,$nm between the two resonances. Including the membrane, we experimentally observed a finesse of $\mathcal{F}_\mathrm{exp}=\finesseExp$ for air-like modes \cite{dam_optimal_2018}, close to the theoretical finesse.
Within air-like modes, we conclude that scattering losses are negligible for one given wavelength. All cavity experiments were performed within the upper right region of \hyperref[fig2]{Fig.\,2(b)}, where the two orientations of SiV$^{\text{-}}$ showed a slight separation in the C-transitions. There, by following the interference fringes, the membrane has an estimated thickness of $\SI{\Membranethicknessimplant}{\micro \meter}$.
Operating at an effective cavity length of  $L_{\mathrm{eff}}\approx \SI{\CavityLength}{\micro\meter}$ the resonator possesses a linewidth of $\kappa/2\pi=\SI{\CavityLinewidth}{\giga\hertz}$, much broader than the Fourier transform limit of a single SiV$^{\text{-}}$ center.
From Gaussian beam optics, we calculated a resonator beam waist of $\omega_0=\SI{\BeamWaist}{\micro\meter}$ and the fundamental cavity mode volume of $V = \ModeVolume\lambda_\mathrm{las}^3$ (\hyperref[sec:appendix:cavity]{Appendix \ref{sec:appendix:cavity}}). Additionally, we determined the quality factor $Q=\QualityFactor$ from  $\mathcal{F}_\mathrm{exp}$, see \hyperref[tab1]{Tab. I} for more details.
This allowed for the estimation of the theoretical Purcell factor
  \begin{equation}
      f_{\mathcal{P}}= \frac{3}{4\pi^2} \left( \frac{\lambda}{n} \right)^3 \frac{Q}{V}
      \label{eq:purcelltheo}
  \end{equation}
which we estimated to $ f_{\mathcal{P}}=\PurcellTheo$.

\subsection*{Cavity-Modulated Photoluminescence and Photoluminescence Excitation}

\begin{figure*}[ht]
  \centering
   \includegraphics[scale=1]{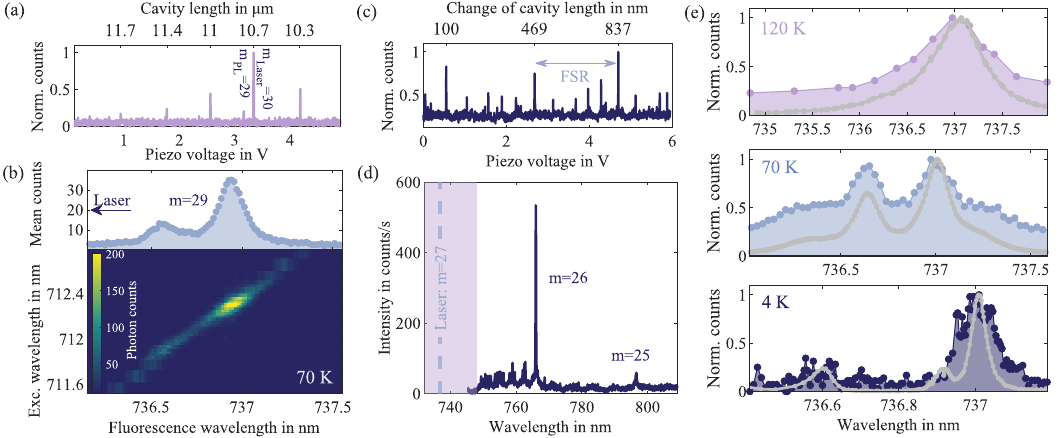}
    \caption{\textbf{
        Cavity-assisted photoluminescence and photoluminescence excitation.}
    \textbf{(a)} Photoluminescence counts from cavity length scan under off-resonant excitation at \SI{712.3}{\nano\meter}. 
    \textbf{(b)} Spectra with cavity-modulated zero phonon line emission (m=29) under off-resonant excitation at \SI{70}{\kelvin}. The excitation laser was resonant with the next higher cavity mode (m=30). The averaged counts from all spectra are shown at the top.  
    \textbf{(c)} Cavity length scan under resonant excitation at \SI{737}{\nano\meter} at \SI{4}{\kelvin}.
    \textbf{(d)} Cavity-modulated sideband emission of the SiV$^{\text{-}}$ emitters at a cavity length of \SI{\CavityLength}{\micro\meter} at \SI{4}{\kelvin}. The excitation laser was resonant with the zero phonon line at \SI{737}{\nano\meter} and the cavity mode m=27.  
    \textbf{(e)} Temperature-dependent photoluminescence excitation, with a sideband collection. Free space data are shown in gray. 
    }
    \label{fig4}
\end{figure*}

Within the cavity, off-resonant excitation was performed with one fundamental transverse mode of the cavity tuned to resonance with the \ac{ZPL} and the next higher mode to resonance with the excitation laser at a lower wavelength of about \SI{712}{\nano\meter}. Cavity length-dependent detection of the SiV$^{\text{-}}$ emitters' \ac{ZPL} in \hyperref[fig4]{Fig.\,4(a)} shows \ac{PL} with a mode volume $V$ between $45$ and $53 \lambda_\mathrm{las}^3$ at \SI{120}{\kelvin}. The highest count rate was observed when the resonator condition $ L_{\mathrm{eff}} =m\frac{\lambda}{2}$ is best fulfilled for the \ac{ZPL} and the laser simultaneously, here with the laser at resonance m=30 and the detection through mode m=29. This translates to an effective cavity length of $L_{\mathrm{eff}}=\SI{10.7}{\micro \meter}$.
Spectral probing of the ZPL emission at \SI{70}{\kelvin} is shown in \hyperref[fig4]{Fig.\,4(b)}, where the cavity resonance is in tune with the excitation laser. 
Adding up the spectra, two peaks become visible. The first results from transitions originating from the upper excited state and the second from the lower excited state. 

For a resonant \ac{PLE} probing, we tuned the laser to the \ac{ZPL} and detected the cavity-modulated phonon sideband. 
In a length scan, we collected fluorescence when the laser was in resonance with the cavity. This reveals three fundamental transverse modes, plus additional higher-order modes as shown in  \hyperref[fig4]{Fig.\,4(c)}.
With the laser in resonance with the cavity (m=27) and the emitters' \ac{ZPL}, we acquired a spectrum shown in \hyperref[fig4]{Fig.\,4(d)}. The sideband couples to the next two cavity modes (m=26, 25). Emission in mode m=25 at $\SI{797}{\nano \meter}$ is still apparent, even with the reflectivity of the mirrors decreasing drastically. 
Resonant \ac{PLE} probing is shown in \hyperref[fig4]{Fig.\,4(e)} for \SI{120}{\kelvin}, \SI{70}{\kelvin} and \SI{4}{\kelvin}. For every excitation wavelength, the cavity length was scanned multiple times while collecting the laser signal and the fluorescence signal. We summed the fluorescence counts, around the cavity resonance for every scan and normalized the data. Comparing the temperature-dependent spectra from within the cavity to the free space \ac{PLE} data, the inhomogeneous linewidth is broader. We attribute this to the cavity mode area being more than 3 times larger than the confocal area resulting in more contributing emitters and a broader spectral distribution. With the larger probe area, we collected the emission from more emitters, including a wider range of strain values, so the emitter ensemble is wider spectrally distributed. At \SI{4}{\kelvin} the C-line shows an inhomogeneous linewidth of $\nu_{\mathrm{inh }}=\SI{\LinewidthinhomoCavfourK}{\giga\hertz}$ ($\SI{\LinewidthinhomoFSfourK}{\giga\hertz}$) within the cavity (free space), a factor of 10 broader than the cavity linewidth $\kappa/2\pi$. 
\ac{PL} plus off-resonant and resonant \ac{PLE} are well-established techniques for confocal microscopy of color centers, which we implemented for the cavity system.
To verify the number of contributing emitters, we compared free space \ac{PLE} counts of the ensemble to those of the single emitter from \hyperref[sec:appendix:single]{Appendix \ref{sec:appendix:single}}. The power-corrected count rates differed by a factor of $\NumberEmitterConfocal$, which also reflects the number of emitters within the confocal mode, consistent with the implantation parameters.
For the larger cavity mode, we expect an order of up to $\NumberEmittersCav$ contributing emitters.

\subsection*{In-Cavity Absorption of Emitter Ensemble}

\begin{figure*}[ht]
  \centering
   \includegraphics[scale=1]{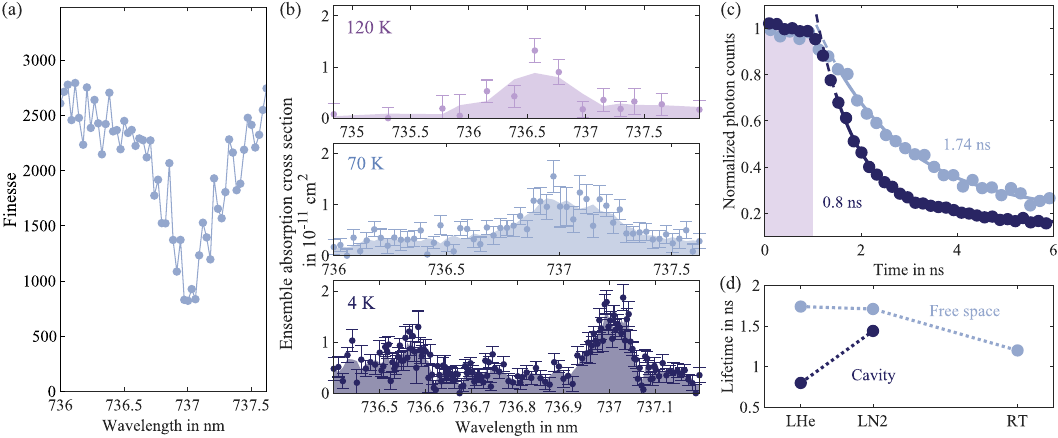}
    \caption{\textbf{
        Absorption spectroscopy and Purcell enhancement of SiV$^{\text{-}}$  transition.}
    \textbf{(a)} Highest observed finesse reduction due to absorption of the emitters.
    \textbf{(b)} Calculated absorption for the emitter ensemble for different temperatures. The cross-section (dots) was estimated from averaged finesse measurements. Error bars show the standard deviation, shaded area shows the moving average for better visibility.  
    \textbf{(c)} Optical lifetime measurement of the C-transition without (light blue dots) and with the cavity (blue dots) with respective exponential decay (solid line) at $\SI{4}{\kelvin}$. The purple area indicates the end of the laser pulse.
    \textbf{(d)} Optical lifetimes at different temperatures without (light blue dots) and with the cavity (blue dots). The dotted lines illustrate the trend.
    }
    \label{fig5}
\end{figure*}

\begin{table*}[ht]
    \centering 
    \caption{\textbf{Membrane, emitter, and cavity parameters.}}
    \label{tab1}
    \begin{tabular}{ l  l l  r  l l }
     \hline  \hline   
\textbf{Parameter} & \textbf{Condition} &  \textbf{Symbol}  &  \textbf{Value}   &\textbf{Unit} &\textbf{Origin} \\ 
\hline     \hline
\textbf{Membrane \& emitter} &   &    & \\ 
Minimal membrane thickness & & $l$ & \Membranethickness & $\si{\nano\meter}$ &  profilometer \\ 
Surface roughness & &  RMS & \Membraneroughness & $\si{\nano\meter}$ &  atomic force microscope \\ 
Implantation depth & & $d$ & \Implantationdepth &$\si{\nano\meter}$  &  implantation energy \\ 
Single emitter linewidth & $\SI{4}{\kelvin}$  &$\nu_{\mathrm{single}}$ & \Emitterlinewidthsingle &$\si{\mega\hertz}$ &  fitted PLE spectrum  \\ 
Inhomogeneous emitter linewidth & $\SI{4}{\kelvin}$  &$\nu_{\mathrm{inh}}$ & \LinewidthinhomoFSfourK &$\si{\giga\hertz}$ &  fitted PLE spectrum \\ 
\hline
\textbf{Mirror} &   &    & \\ 
Radius of curvature x &  &RoC$_\mathrm{x}$ & \ROCx & $\si{\micro\meter}$   & fitted interferometer data  \\ 
Radius of curvature y &  &RoC$_\mathrm{y}$ & \ROCy & $\si{\micro\meter}$    & fitted interferometer data  \\ 
Mirror reflectivity &   $\SI{737}{\nano\meter}$ & R & \MirrorReflectivity & \%  & reflection data from Laseroptik\\  
Transmission &   $\SI{737}{\nano\meter}$ & $T_{\mathrm{ppm}}$ & \MirrorTransppm & ppm   & reflection data from Laseroptik\\   
\hline
\textbf{Cavity} &   &    & \\ 
Theoretical finesse & & $\mathcal{F}_{\mathrm{theo}}$ & \finesseTheo &  &  $T_{\mathrm{ppm}}$ \\ 
Experimental finesse & & $\mathcal{F}_{\mathrm{exp}}$ & \finesseExp &  &  cavity length scan \\ 
Typical cavity length & & $L_\mathrm{eff}$  &\CavityLength  &$\si{\micro\meter}$ & calculation from spectra \\ 
Typical cavity linewidth&  &$\kappa$ & \CavityLinewidth &$\si{\giga\hertz}$ \ \ &   $\mathcal{F}_{\mathrm{exp}}$ and $L_\mathrm{eff}$\\ 
Beam waist & &$\omega_0$  & \BeamWaist & $\si{\micro\meter}$ &  $L_\mathrm{eff}$ and RoC \\
Typical mode volume & & $V$  & \ModeVolume &$\lambda_\mathrm{las}^3$  &  $\omega_0$ \\ 
Typical mode number&  &$m$ & \ModeNumber &  & calculation from spectra \\  
Quality factor&  &$Q$ & \QualityFactor & &  $\mathcal{F}_{\mathrm{exp}}$ and  $m$ \\ 
\hline 
\textbf{Purcell enhancement} &   &    & \\ 
Theoretical Purcell factor &  &$f_{\mathcal{P}}$ & \PurcellTheo & & equation (\ref{eq:purcelltheo})  \\ 
Cavity-emitter lifetime &  $\SI{70}{\kelvin}$ &$\tau_{\mathrm{cav}}$ & \LifetimeEmitterCavitySev & ns   & fitted lifetime data  \\ 
Cavity-emitter lifetime &  $\SI{4}{\kelvin}$ &$\tau_{\mathrm{cav}}$ & \LifetimeEmitterCavity & ns   & fitted lifetime data  \\ 
Effective Purcell factor &  $\SI{4}{\kelvin}$ & $F_{P\mathrm{,eff}}$ & \PurcellExp &   &   equation (\ref{eq:purcellExperiment}) \\ 
Cooperativity&  $\SI{4}{\kelvin}$  & $C$ & \Cooperativity &  &   equation (\ref{eq:coop})  \\ 
\hline \hline    
     \end{tabular}
\end{table*}

Quantum communication protocols profit from a direct way of resonantly manipulating and reading quantum states. Reflection-based approaches in nanophotonic cavities presented the crucial probing of a system by resonant excitation and detection \cite{nguyen_integrated_2019, nguyen_quantum_2019, bhaskar_experimental_2020,stas_robust_2022}. 
When we resonantly probed the emitters' emission out of the cavity, we also collected and separated the reflected laser signal by optical filtering. This provided information on the cavity in-coupling and scattering within the resonator. 
We observed less prominent cavity resonances within an air-like mode when the system is resonant with the emitter ensemble. Probing the finesse, by scanning the cavity over two cavity resonances, showed reduced finesse values for the SiV$^{\text{-}}$ wavelengths. 
\hyperref[fig5]{Fig.\,5(a)} depicts such measurement, with the highest observed finesse reduction at \SI{70}{\kelvin}. The SiV$^{\text{-}}$ ensemble inside the resonator absorbed the laser light, revealing the emitters' population.
The emitter-cavity interaction directly imprints on the reflected laser signal.
Absorption losses $L_{\mathrm{SiV}}$ introduced by the emitters can be calculated with \cite{hausler_diamond_2019}
\begin{equation}
    L_{\mathrm{SiV}}=\pi \left( \frac{1}{\mathcal{F}_{\mathrm{SiV}}} -\frac{1}{\mathcal{F}_{\mathrm{exp}}} \right) \ .
\end{equation}
$\mathcal{F}_{\mathrm{exp}}$ describes the finesse of the membrane-cavity system and $\mathcal{F}_{\mathrm{SiV}}$ the emitter-reduced finesse.
Considering the maximum reduction of the finesse down to $\mathcal{F}_{\mathrm{max}}\approx820$ at the C-transition, the emitters together introduced losses of \LossesMaxreduction\,ppm. 
Combining the losses with the area of the resonator mode 
$A=\SI{\CavityModeArea}{\micro\meter}^2$
the absorption cross-section of the emitters can be obtained:
\begin{equation}
    \sigma_{\mathrm{abs}}^{\mathrm{ens}}= L_{\mathrm{SiV}} \cdot A \ .
\end{equation}
We calculated an absorption cross section of $\sigma_{\mathrm{abs}}^{\mathrm{ens}}=\sigmamaxreduction\,\mathrm{cm}^2$ for the ensemble.
For the number of contributing emitters N, the absorption cross-section of a single emitter excluding nonlinear effects \cite{hausler_diamond_2019} follows:
\begin{equation}
    \sigma_{\mathrm{abs}}^{\mathrm{single}}= \frac{\sigma_{\mathrm{abs}}^{\mathrm{ens}}}{N} \ .
\end{equation}
With approximately $\NumberEmittersCav$ emitters contributing, we calculated a cross-section of 
 $\sigma_{\mathrm{abs}}^{\mathrm{single}}\approx\sigmamaxreductionsingle\,\mathrm{cm}^2$ for a single emitter, which falls within the middle range of reported values \cite{neu_photophysics_2012,hausler_diamond_2019}.
Furthermore, we calculated the absorption cross-section of the emitter-cavity system (\hyperref[fig5]{Fig.\,5(b)}) from multiple cavity resonances for $T=(120, 70, 4)\,$K, respectively. 
Comparing the highest ensemble absorption cross-section $\sigma_{\mathrm{abs}}^{\mathrm{ens}}=(\sigmawarmens, \sigmamiddleens, \sigmacoldens)\,  \cdot 10^{-11}\,\mathrm{cm}^2$ for these temperatures, the absorption enhances as the inhomogeneous linewidth of the emitters decreases with lower temperatures.
Overall, absorption of the emitters can be determined due to good coupling of the emitter ensemble to the cavity mode. For a quantification of the enhancement, we performed ordinary lifetime measurements. 

\subsection*{Purcell Enhancement}

Experimental evidence for Purcell-enhanced lifetime-shortening is the direct comparison of the optical lifetime of the emitters in free space ($\tau_{\mathrm{fs}}$) with the lifetime inside the cavity field ($\tau_{\mathrm{cav}}$):
   \begin{equation}
    F_{P\mathrm{,eff}}= \frac{\tau_{\mathrm{fs}}}{\tau_{\mathrm{cav}}} =1+\xi f_{\mathcal{P}} \ .
      \label{eq:purcellExperiment}
  \end{equation}
With the branching ratio $\xi$ the theoretical Purcell factor from equation (\ref{eq:purcelltheo}) can be connected with the effective Purcell factor $F_{P\mathrm{,eff}}$.
To determine the effective Purcell factor we measured the lifetime of the emitter ensemble at $\SI{4}{\kelvin}$ resonantly with the C-transition within the cavity and in free space under off-resonant excitation.
The fitted lifetime data in \hyperref[fig5]{Fig.\,5(c)} verify a shortening of the optical lifetime from $\tau_{\mathrm{fs}}=\SI{\LifetimeEmitterFS}{\nano \second }$ to $\tau_{\mathrm{cav}}=\SI{\LifetimeEmitterCavity}{\nano \second }$. 
 This results in a Purcell factor of $F_{P\mathrm{,eff}}=\PurcellExp$ at liquid helium temperatures (LHe).
The free space lifetime is in good agreement with previous studies on the SiV$^{\text{-}}$ color center \cite{rogers_multiple_2014,jahnke_electronphonon_2015,marseglia_bright_2018}. Under ambient conditions the free space lifetime was shorter, see \hyperref[fig5]{Fig.\,5(d)}. Again, this agrees with the studies and is attributed to the thermal activation of a non-radiative decay channel resulting in a poor quantum yield \cite{rogers_multiple_2014}. 
The measured free space optical lifetimes of the emitter ensemble and previous studies can be found in \hyperref[tab2]{Tab. II} in \hyperref[sec:appendix:lifetime]{Appendix \ref{sec:appendix:lifetime}}.
It follows that at liquid nitrogen temperature (LN2) the Purcell factor decreases ($F_{P\mathrm{,eff}}=1.2$), as the linewidth of the emitter ensemble broadens towards room temperature (RT), due to a reduced overlap of cavity and emitters' linewidth.
The clear sign of the Purcell effect at $\SI{4}{\kelvin}$, which enabled for higher emission rates of the strained emitters, allowed the estimation of important cavity parameters. 
These include the beta-factor
\begin{equation}
    \beta=\frac{F_{P\mathrm{,eff}}-1}{F_{P\mathrm{,eff}}} \ ,
\end{equation}
of $\beta=\betafactor$
and the cooperativity
\begin{equation}
    C=F_{P\mathrm{,eff}}-1
      \label{eq:coop}
\end{equation}
of above 1, towards high cooperativity ($C$>>1).

\section*{Conclusion}

In this work, we presented a diamond-cavity interface based on an ensemble of SiV$^{\text{-}}$ centers coupled to the mode of an open Fabry-Perot resonator. The SiV$^{\text{-}}$ ensemble created by ion implantation into a µm-thin diamond membrane possesses a wide range of laterally continuously varying strain values. The induced strain causes significant shifts of the fine-structure transitions, which differ for the different orientations of the SiV$^{\text{-}}$ classes in the ensemble. The ground-state splitting $\Delta_{\mathrm{gs}}$ reaches values of up to $\SI{\groundstatesplitmax}{\giga \hertz}$ across the diamond membrane. The strain can be used to overlap the optical transitions of different classes of SiV$^{\text{-}}$ centers, as observed by a crossing in the corresponding spectral lines. Furthermore, the large ground-state splitting is highly important for the future use of the system as a spin-photon interface at moderate temperatures above 1\,Kelvin. Thus, a fully tunable microcavity enables to selective choose specifically strained emitters. While lateral tunability allows the selection of an appropriate strain value, tunability of the resonance frequency enables deterministic coupling of individual transitions to the cavity mode. Cavity-assisted absorption yields a cross section of $\sigma_{\mathrm{abs}}^{\mathrm{ens}}=\sigmamaxreduction\,\mathrm{cm}^2$ of the SiV emitter ensemble resulting in $\sigma_{\mathrm{abs}}^{\mathrm{single}}\approx\sigmamaxreductionsingle\,\mathrm{cm}^2$ for a single SiV$^{\text{-}}$ color center. The lifetime shortening of the emitter ensemble in the sub-ns regime revealed a Purcell factor of $F_{P\mathrm{,eff}}=\PurcellExp$. The fourfold deviation from the theoretical predictions can be connected to the branching ratio.
The resulting cooperativity of $C=1.2$ yields the cavity quantum electrodynamics parameter set of $(g,\kappa,\gamma)/2\pi = (\gcavity,\CavityLinewidth,\gammacavity)\ $GHz.



\section*{Funding}
 The authors gratefully acknowledge the German Federal Ministry of Education and Research (BMBF) funding within the project QR.X (16KISQ005, 16KISQ006). The authors gratefully acknowledge the funding from the European Union and the DFG within the Quantera-project SensExtreme. S.S. acknowledges funding from the Marie Curie ITN project LasIonDef (GA n.956387).

\section*{Acknowledgments}
 We thank Richard Waltrich and Susanne Menzel for their support in the cleaning process, Patrick Maier and Jan Schimmel for cavity structure fabrication, and Johannes Lang and Jens Fuhrmann for the annealing of the diamond sample. Most measurements were performed with the QuDi software suite \cite{binder_qudi_2017}.

\section*{Disclosures}
The authors declare no conflicts of interest.

\section*{Data Availability Statement}
Data underlying the results presented in this paper are not publicly available at this time but may be obtained from the authors upon reasonable request.

\section*{Appendix}

\subsection{Diamond Membrane Thickness}
\label{sec:appendix:thickness}

\begin{figure}[htbp]
     \centering
   \includegraphics[scale=1]{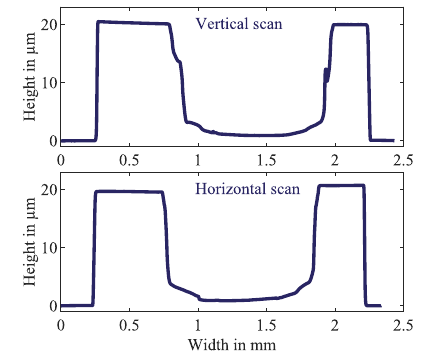}
    \caption{\textbf{
       Height variation of the bonded diamond membrane along the main axes.}
       Horizontal (left to right) and vertical (bottom to top) scans with a profilometer, performed after all optical measurements, started and ended at the mirror surface. The dimensions of the diamond sample become apparent with the etched window of approximately $\SI{1}{\milli \meter}\times\SI{1}{\milli \meter}$.
    }
    \label{figS_fab}
\end{figure}

Scans taken with a profilometer (Dektak, Bruker Corp.) along the two axes of the \ac{DM} show the thickness variations along the whole membrane, see \hyperref[figS_fab]{Fig.\,6}.
The minimum thickness in the two scans is $\SI{850}{\nano\meter}$ and $\SI{880}{\nano\meter}$ respectively, while the frame, used for handling the membrane, shows a thickness of $\SI{20}{\micro\meter}$. 

\subsection{Implantation and Cleaning}
\label{sec:appendix:fab}

\begin{figure}[htbp]
  \centering
   \includegraphics[scale=1]{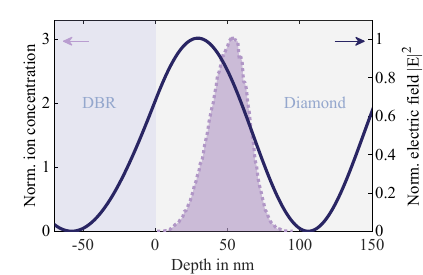}
    \caption{\textbf{Diamond-mirror interface with electric field distribution and implantation depth inside the membrane. } 
The electric field (blue line) for \SI{737}{\nano \meter} within the DBR (blue area) and the bonded membrane (gray area) from a transfer-matrix simulation, using the coating design, is shown.
The implantation depth (purple, dashed line) from a SRIM simulation, indicates the overlap of the electric field with the emitters.
    }
    \label{figS_implant}
\end{figure}

An implantation depth inside the \ac{DM} for the bonded side was calculated from Stopping Range of Ions in Matter (SRIM) simulation. 
\hyperref[figS_implant]{Fig.\,7} shows the simulated depth distribution of the ion concentration together with the estimated electric field inside the single crystal membrane. The electric field simulation for the diamond-mirror interface is based on a transfer-matrix, using the DBR coating design, and the assumption of the diamond bonded without an air gap. 
The highest ion concentration lies at 80\% of the highest field intensity for \SI{737}{\nano \meter} inside the diamond. 
The ion concentration allows for an estimation on the emitter creation yield. The creation of emitters in diamond by thermal annealing is generally of low efficiency \cite{luhmann_coulomb-driven_2019}.
Different values have been reported for the yield of SiV$^{\text{-}}$, ranging from 2.5\,\% to 18\,\% \cite{tamura_array_2014,schroder_scalable_2017,lagomarsino_optical_2018,zuber_shallow_2023}.

After implantation and annealing, the \ac{DM} was cleaned and bonded three times. While the first bonding attempt with RT tri-acid cleaning resulted in reduced finesse values due to scattering losses, we performed a second attempt with standard tri-acid cleaning ($130^{\circ}$C). This resulted in less loss, although still observable. The third approach involved hydrofluoric acid cleaning followed by tri-acid cleaning ($130^{\circ}$C), which resulted in a strong bond and negligible scattering losses within air-like modes. For bonding, the membrane was directly transferred from ultrapure water droplets onto the flat mirror.

\subsection{Confocal Setup and Cavity Characteristics}
\label{sec:appendix:cavity}

For excitation, we used the laser light of an actively stabilized titanium-sapphire ring laser (Matisse 2 TS, Sirah 
Lasertechnik GmbH), power-stabilized with an acoustic optical modulator (AOM, 3350-199, Gooch \& Housego Ltd.). 
For pulsed measurements, we employed a fiber-pigtailed amplitude electro-optical modulator (EOM, AM705, JENOPTIK AG) regulated by either a fast rise-time pulse generator (\SI{30}{\pico \second} rise/fall time; LBE-1320, Leo Bodnar Electronics Ltd.) or an arbitrary waveform generator (AWG, M9502A, Keysight Technologies Inc.). 
The excitation laser was guided into the cryostat via a single-mode fiber (SM CU600PSC, IVG Fiber Ltd.) and then partially reflected by a wedge prism (0.5° Nom. $\SI{12.5}{\milli\meter}$ Dia. VIS-NIR, N-BK7 Wedge Prism, Edmund Optics) to the aspherical lens. 
For detection, the reflected laser and fluorescence signal out of the cavity was collected with a multimode fiber (MM FG050LGA, Thorlabs Inc.) and guided out of the cryostat. It was separated by spectral filtering: 730 notch filter (single notch filter 730\,nm - zet 730NF, Chroma Technology Corp.), 740/13 bandpass filter (740/13 BrightLine HC, Semrock), 750 longpass filter (Thorlabs Inc. and AHF). The notch filter blocked the laser or unwanted fluorescence around the \ac{ZPL}, the bandpass filtered the \ac{ZPL}, and the longpass separated the laser signal from the phonon sideband. 
Laser light was sent to a silicon avalanche photodetector (APD120A2/M, Thorlabs Inc.), and fluorescence signal either to a single photon counting module (SPCMAQRH-14, Excelitas Technologies Corp.) or a spectrometer (SpectraPro HRS-500, Teledyne Princeton Instruments).
Correlation measurements were conducted with a correlation device (Time Tagger Ultra module, Swabian Instruments GmbH). 
In transmission, the laser was directly detected with a silicon photodiode (S1337-66BQ, Hamamatsu Photonics K.K.) placed below the planar mirror. 

\begin{figure}[htbp]
  \centering
    \includegraphics[scale=1]{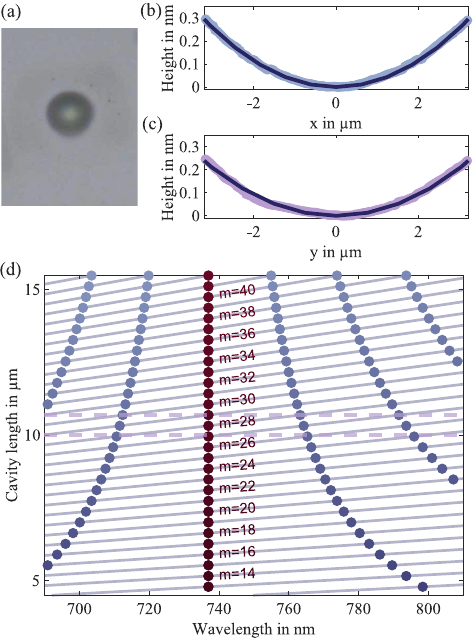}
    \caption{\textbf{
    Spherical structure of the hemispherical cavity and simulated mode dispersion.}
    \textbf{(a)} Microscope image of the concave spherical mirror structure from CO$_2$-laser ablation.
    \textbf{(b)} Height profile (light blue) of the concave structure along the most curved axis reveals a \ac{RoC} of $\SI{\ROCx}{\micro\meter}$.
    \textbf{(c)} Height profile (purple) of the concave structure along the least curved axis reveals a \ac{RoC} of $\SI{\ROCy}{\micro\meter}$.
    \textbf{(d)} The simulated cavity resonance condition is shown for different cavity lengths and wavelengths (blue lines). For a cavity resonant at \SI{737}{\nano\meter} (red dots), the corresponding lower and higher resonance positions (blue dots) are highlighted. The effective cavity lengths used in the measurement are indicated by the horizontal lines (purple).
    }
      \label{figS_params}
\end{figure}

Characterization of the spherical cavity structure (\hyperref[figS_params]{Fig.\,8(a)}) in a home-built Twyman-Green-interferometer revealed the \ac{RoC} and the ellipticity, specified in the main text.
\hyperref[figS_params]{Fig.\,8(b)} and \hyperref[figS_params]{8(c)} show the height profile from the interferometric analysis along the two vertical axes of the structure. For reference, spheres with the corresponding \ac{RoC} from the fit are shown on top (dark blue).
The \ac{RoC} allowed for the calculation of the beam waist within the resonator
\begin{equation}
    \omega_0 \approx \sqrt{\frac{\lambda}{\pi}} \left( L_{\mathrm{eff}} RoC-L_{\mathrm{eff}}^2\right)^{1/4} \ ,
\end{equation} 
used for the estimation of the fundamental cavity mode volume $V=\frac{\pi}{4} L_{\mathrm{eff}}\omega_0^2 $ and the cavity mode area $A=\frac{\pi\omega_0^2}{4}$.

Estimation on the cavity lengths are based on the position of cavity resonances in simulation \hyperref[figS_params]{Fig.\,8(d)}. The free spectral range increases for shorter cavity lengths. Effective cavity lengths from off-resonant excitation with the ZPL resonant with m=29 (\hyperref[fig4]{Fig.\,4(b)}) and resonant excitation with the ZPL resonant with m=27 (\hyperref[fig4]{Fig.\,4(d)}) are highlighted. 

\subsection{Single SiV Emitter}
\label{sec:appendix:single}

\begin{figure}[htbp]
  \centering
   \includegraphics[scale=1]{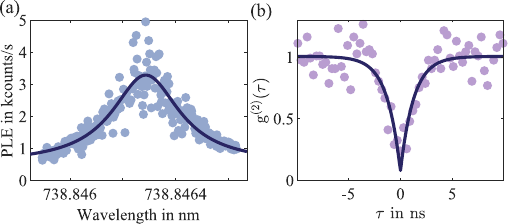}
    \caption{\textbf{
        Single SiV$^{\text{-}}$ color center in a diamond membrane.}
    \textbf{(a)} Photoluminescence excitation of a SiV$^{\text{-}}$ emitter with a center wavelength of $\SI{738.8463}{\nano \meter}$.
    \textbf{(b)} Correlation function $g^{2}(\tau)$ of the SiV$^{\text{-}}$ emitter from (a) in a Hanbury-Brown-Twiss experiment. The photon antibunching suggests the emission of a single SiV$^{\text{-}}$ emitter inside the \ac{DM}.
    }
    \label{figS_single}
\end{figure}

Under resonant excitation, the membrane showed emission from single SiV$^{\text{-}}$ color centers for longer wavelengths compared to the emitter ensemble within the implanted area. A \ac{PLE} scan of a single emitter above $\SI{738.8}{\nano \meter}$ in confocal mode is shown in \hyperref[figS_single]{Fig.\,9(a)}. At a low excitation power of $\SI{\powerPLEsingle}{\nano\watt}$ the emitter shows a narrow linewidth of $\SI{\Emitterlinewidthsingle}{\mega\hertz}$ close to the Fourier transform limit.
Single emitter emission verified with a second order correlation $g^2(\tau)$ measurement in a Hanburry-Brown-Twiss configuration with \SI{550}{\nano\watt} resonant laser yields a dip of $g^2(0)= 0.1$ (\hyperref[figS_single]{Fig.\,9(b)}). This suggests almost pure emission of one SiV$^{\text{-}}$ color center in a resonant drive. 

\subsection{Optical Lifetime}
\label{sec:appendix:lifetime}

The optical lifetime of SiV$^{\text{-}}$ emitters decreases from low temperature to RT due to other non-radiative decay channels. 
Free space optical lifetime measurements for SiV$^{\text{-}}$ in diamond are summarized in \hyperref[tab2]{Tab. II} for ambient and low-temperature conditions. Two reported values are from bulk samples, one from nanowires, although the authors did not attribute any Purcell enhancement to their structures. For comparison, the measured values with off-resonant pulsing by a $\SI{700}{\nano \meter}$ diode laser (Taiko PDL M1 and DH-IB-705-B, PicoQuant GmbH) for the single-crystal of this work are shown. The values are in good agreement with the reported lifetimes.

\begin{table}[htbp]
    \label{tab2}
    \caption{\textbf{Reported optical lifetimes for SiV$^{\text{-}}$ in ns at different temperatures.} 
    Amongst them are lifetimes of SiV$^{\text{-}}$ in bulk diamond: 
    Rogers et al. \cite{rogers_multiple_2014} and
    Jahnke et al. \cite{jahnke_electronphonon_2015}, in nanowires:  Marseglia et al. \cite{marseglia_bright_2018}, and in a membrane: this work (free space).}
    \begin{tabular}{ l l l l l l   }
     \hline  \hline   
& & \textbf{Bulk} &  \textbf{Bulk}  &  \textbf{Nanowires} &  \textbf{\ac{DM}}  \\ 
& & \cite{rogers_multiple_2014} & \cite{jahnke_electronphonon_2015} & \cite{marseglia_bright_2018} & This work    \\ 
\hline     \hline
\textbf{RT}   &             & 1.28 & 1.1 & 1.22  &  \LifetimeEmitterFSRT \\ 
\textbf{70\,K} \ \ & & - & 1.54& - & \LifetimeEmitterFSSeventyK \\ 
 \textbf{4\,K} & & 1.72 & 1.55 & 1.73 & \LifetimeEmitterFS \\
\hline \hline    
     \end{tabular}
\end{table}

\FloatBarrier 
\bibliography{article_main} 

\end{document}